\RequirePackage{fix-cm}
\documentclass[smallextended]{svjour3}       
\smartqed  
\usepackage{graphicx}
\begin{document}

\title{Relating  quantum coherence and correlations with entropy-based measures
}


\author{Xiao-Li Wang   \and Qiu-Ling Yue  \and Chao-Hua Yu \and Fei Gao\and Su-Juan Qin 
}


\institute{X.-L. Wang  \and Q.-L. Yue \and C.-H. Yu  \and
        F. Gao \and S.-J. Qin \at
              State Key Laboratory of Networking and Switching Technology, Beijing University of Posts and Telecommunications, Beijing 100876, China \\
              \email{qsujuan@bupt.edu.cn}           
           \and
           X.-L. Wang  \at
              School of Mathematics and Information Science, Henan Polytechnic University, Jiaozuo, 454000, China\\
              State Key Laboratory of Information Security (Institute of Information Engineering, Chinese Academy of Sciences, Beijing 100093)
}

\date{Received: date / Accepted: date}

\maketitle

\begin{abstract}
  Quantum coherence as an important quantum resource plays a key role in quantum theory. In this paper, using entropy-based measures, we investigate the relations between quantum correlated coherence, which is the coherence between subsystems [K. C. Tan, H. Kwon, C. Y. Park, and H. Jeong, Phys. Rev. A \textbf{94}, 022329 (2016)], and two main kinds of quantum correlations as defined by quantum discord as well as quantum entanglement. In particular, we show that quantum discord and quantum entanglement can be well characterized by quantum correlated coherence. Moreover, we prove that the entanglement measure formulated by quantum correlated coherence is lower and upper bounded by the relative entropy of entanglement and the entanglement of formation, respectively, and equal to the relative entropy of entanglement for maximally correlated states.

\end{abstract}

\section{Introduction}
 Quantum coherence  arising from quantum superposition [1],  represents a fundamental feature that marks the departure of quantum mechanics from classical physics. Recently, many efforts have been devoted to develop the resource theory of quantum coherence [2-7]. Meanwhile, various properties of quantum coherence have been investigated such as the connections between quantum coherence and quantum correlations in multipartite systems [8-13], the distillation of coherence [5, 14, 15], the dynamics under noisy evolution of quantum coherence [16, 17], among others. The role of coherence in thermodynamics [18, 19] has also been discussed.

  Quantum coherence in multipartite systems involves the essence of quantum correlations. One of the potential quantum correlations is quantum entanglement [20-24] which has been widely concerned. Another kind of quantum correlations is quantum discord [25-29], which may even exist in a separable state with vanished entanglement. Both of them are crucial resources for the development of quantum technologies, such as quantum communication [30, 31], quantum computation [32, 33], quantum metrology [34], and many more. In these cases, quantum correlations indicate an advantage of quantum methods over classical ones.

 Note that previous results in Ref. [13] have established a unified view of quantum discord and quantum entanglement with the framework of quantum coherence based on the $l_{1}-$norm of coherence. By contrast, we will adopt the relative entropy of coherence to explore the concise relations  between quantum discord and quantum correlated coherence [13], which is the coherence between subsystems. In fact, quantum correlated coherence is a `correlation function', which captures the correlation between subsystems. Besides, quantum correlated coherence can  be used to construct an entanglement measure, which is called the entanglement of coherence (EOC) [13]. However, Many important properties, such as additivity, and relations to other entanglement measures, have not been  investigated. In this paper, using entropy-based measures, we will show that the EOC is lower and upper bounded by the relative entropy of entanglement and the entanglement of formation, and equal to the relative entropy of entanglement for maximally correlated states. We also compare the EOC with the entanglement measure which is the minimal discord over state extensions [35]. Our work provides clear relations between  quantum coherence and correlations with entropy-based measures.

The rest of this paper is organized as follows. Sec. 2 introduces some definitions of quantum coherence, quantum entanglement and quantum discord. In Sec. 3, we give the relations between quantum  correlated coherence and quantum discord. In Sec.4, we prove the bounds of the entanglement of coherence (EOC), which is formulated by quantum correlated coherence with respect to the relative entropy of coherence. This paper is ended with the conclusion in Sec. 5.

\section{Preliminaries}
  In the framework of coherence theory introduced in Ref. [4], let $\{|i\rangle\}$ be a fixed basis in the finite dimensional Hilbert space, and the incoherent sates  are those whose density matrices are diagonal in this fixed basis, being of the form $\sum_{i}p_{i}|i\rangle\langle i|$  where $p_{i}$ are probabilities. The set of all incoherent states is denoted by $\mathcal{I}$. It is known that quantum operations are characterized by a set of Kraus operators $\{K_{l}\}$  satisfying $\sum_{l}K^{\dag}_{l}K_{l}=I$. In particular, an incoherent  quantum operation is that for which there exists a Kraus representation $\{K_{l}\}$  such that$ \frac{K_{l}\sigma K^{\dag}_{l}}{Tr(K_{l}\sigma K^{\dag}_{l})}\in \mathcal{I}$ for all $l$ and all $\sigma \in \mathcal{I}$. The von Neumann measurement with respect to the  fixed basis $\{|i\rangle\}$ (otherwise called the dephasing operation) is a special  incoherent quantum operation  denoted by $\mathrm{\Pi}= \{|i\rangle \langle i|\}$. For any state $\rho$, we have
$\mathrm{\Pi} (\rho)=\rho ^{diag} =\sum _{i} |i\rangle \langle i| \rho |i\rangle \langle i|$. Remarkably, any state $\rho$ will generate an incoherent state
 $\rho ^{diag}$ by removing all off-diagonal terms from its density matrix in the fixed basis  through the von Neumann measurement  $\mathrm{\Pi}$. In this paper, we will employ the relative entropy of coherence as the coherence measure which is defined as  $\mathcal{C}_{re}(\rho)= \min_{\sigma\in \mathcal{I}}S(\rho\|\sigma)$,  where  $S(\rho\|\sigma)=Tr(\rho\log_{2}\rho)-Tr(\rho\log_{2}\sigma)$ is the quantum relative entropy [36] and the minimization is taken over the set of incoherent
states $\mathcal{I}$. It has been shown that $\mathcal{C}_{re}(\cdot)$ satisfies all the conditions  mentioned in Ref. [4]. Crucially, this quantity can be evaluated exactly: $\mathcal{C}_{re}(\rho)= S(\rho^{diag})-S(\rho)$, where $S(\rho)=-Tr(\rho \log_{2}\rho)$ is the von Neumann entropy [36].

In this paper, unless otherwise stated, we will often refer to a bipartite system denoted by $AB$, where $A$ and $B$ are local subsystems. For convenience, we say the subsystems $A$ and $B$ are held by Alice and Bob, respectively. For a given state $\rho_{AB}$ in system $AB$, the local states of Alice and Bob are denoted by $\rho_{A}=Tr_{B}(\rho_{AB})$ and $\rho_{B}=Tr_{A}(\rho_{AB})$, respectively, which are obtained by performing a partial trace on $\rho_{AB}$.

Quantum entanglement [20-24] is a popular kind of quantum correlations which can not be prepared by local operations and classical communication (LOCC). The states prepared by LOCC are called separable states which can be represented as a convex combination of product states, i.e.,
$\sigma_{AB}=\sum_{k}p_{k}|\varphi_{k}\rangle_{A}\langle\varphi_{k}|\bigotimes|\phi_{k}\rangle_{B}\langle\phi_{k}|$,
where $\{|\varphi_{k}\rangle_{A}\}$ and $\{|\phi_{k}\rangle_{B}\}$ are normalized but not necessarily orthogonal.  Here, we will refer to the relative entropy of entanglement defined as $E_{re}(\rho_{AB})= \min_{\sigma \in \mathcal{S}} S(\rho\|\sigma)$ with the minimization over the set of separable states $\mathcal{S}$ [21,  22]. Another closely related quantity is  entanglement of formation defined as $E_{f}(\rho_{AB})= \min_{\{p_{k},|\psi_{k}\rangle\}}\sum_{k}p_{k}E_{re}(|\psi_{k}\rangle\langle\psi_{k}|),$ where the minimization is taken over all decompositions of the state  $\rho_{AB}=\sum_{k}p_{k}|\psi_{k}\rangle\langle\psi_{k}|$ [20].

Quantum discord  measures the disturbance induced by local operations to multipartite states [25-29]. Let $\{|i\rangle_{A}\}$ and $\{|j\rangle_{B}\}$ be  orthonormal  bases of subsystems $A$ and $B$, respectively. If Bob performs the von Neumann measurement $\mathrm{\Pi}_{B}= \{|j\rangle _{B}\langle j|\}$ on his subsystem, the post-measurement state is denoted as
\begin{equation}
\mathrm{\Pi} _{B}(\rho_{AB})=\sum _{j} (I_{A}\bigotimes|j\rangle_{B} \langle j| )\rho_{AB} (I_{A}\bigotimes|j\rangle_{B} \langle j| ).
\end{equation}
The asymmetric quantum discord with respect to  $\mathrm{\Pi}_{B}$ can be written in terms of a difference of relative entropies [25-27],
\begin{equation}
D_{\mathrm{\Pi}_{B}}^{A|B}(\rho_{AB})= S(\rho_{AB}\|\rho_{A}\bigotimes\rho_{B})-S\big(\mathrm{\Pi}_{B}(\rho_{AB})\|\rho_{A}\bigotimes\mathrm{\Pi}_{B}(
\rho_{B})\big).
\end{equation}
In the classical-quantum dichotomy, the asymmetric quantum discord is defined as $D^{A|B}(\rho_{AB})= \min_{\mathrm{\Pi}_{B}}(\rho_{AB})$ with the minimization over all local von Neumann measurements $\mathrm{\Pi}_{B}$ to remove the measurement-basis dependence. If Alice only performs the von Neumann measurement $\mathrm{\Pi}_{A}= \{|i\rangle _{A}\langle i|\}$ on her subsystem, the similar results are  available. If both Alice and Bob perform local von neumannn measurements $\mathrm{\Pi}_{A}$ and $\mathrm{\Pi}_{B}$  on their respective subsystems, the symmetric quantum discord (global quantum discord in bipartite system [27]) with respect to $\mathrm{\Pi}_{A}\bigotimes\mathrm{\Pi}_{B}$ is defined as:
$$ D_{\mathrm{\Pi}_{A}\bigotimes\mathrm{\Pi}_{B}}(\rho_{AB})= S(\rho_{AB}\|\rho_{A}\bigotimes\rho_{B})-S\big(\mathrm{\Pi}_{A}\bigotimes\mathrm{\Pi}_{B}(\rho_{AB})\|\mathrm{\Pi}_{A}(\rho_{A})\bigotimes\mathrm{\Pi}_{B}(
\rho_{B})\big).$$
Then, the standard form of symmetric quantum discord is defined as $D(\rho_{AB})= \min_{\mathrm{\Pi}_{A}\bigotimes\mathrm{\Pi}_{B}}(\rho_{AB})$,  with the minimization over all the local von Neumann measurements of Alice and Bob.

 Recently, Yadin et al. [37] have studied the asymmetric basis-dependent discord $D_{\mathrm{\Pi}_{B}}^{A|B}(\cdot)$ which can be seen as the basis-dependent measure of quantumness of  correlation, and the properties of $D_{\mathrm{\Pi}_{B}}^{A|B}(\cdot)$ under the strictly incoherent operations are investigated. Here, we will relate the basis-dependent discord and quantum correlated coherence.
\section{Quantum correlated coherence and quantum discord }
  Let $\{|i\rangle_{A}\}$ and $\{|j\rangle_{B}\}$ be the fixed  bases of subsystems $A$ and $B$ respectively, and we usually use  their tensor product $\{|ij\rangle_{AB}\}$ as the fixed basis of the composite system $AB$. For a state $\rho_{AB}$ in system $AB$, its total coherence  is $\mathcal{C}_{re}(\rho_{AB})$,  while $\mathcal{C}_{re}(\rho_{A})$ and $\mathcal{C}_{re}(\rho_{B})$ are known as local coherences. Whenever $\rho_{AB}$ is a product state, the sum of local coherences is equal to  the total coherence. Generally, the relative entropy of coherence admits the  super-additive property [11]
 \begin{equation}
\mathcal{C}_{re}(\rho_{AB})\geq \mathcal{C}_{re}(\rho_{A})+\mathcal{C}_{re}(\rho_{B}).
\end{equation}
Thus, the definition of quantum correlated coherence with respect to the relative entropy of coherence is given as the following.

\textbf{Definition 1}. (K. C. Tan et al. [13]) Let $\{|i\rangle_{A}\}$ and $\{|j\rangle_{B}\}$ be the fixed bases of subsystems $A$ and $B$ respectively. For a given state $\rho_{AB}$ in system $AB$, its quantum correlated coherence is defined as
 \begin{equation}
\mathcal{C}_{re}^{cc}(\rho_{AB})\equiv \mathcal{C}_{re}(\rho_{AB})-\mathcal{C}_{re}(\rho_{A})-\mathcal{C}_{re}(\rho_{B}).
\end{equation}

Obviously, quantum correlated coherence is the total coherence between subsystems and non-negative. In fact, quantum correlated coherence is a `correlation function' which is similar as quantum mutual information [36]. For arbitrary fixed bases of subsystems $A$ and $B$, the quantum correlated coherence of $\rho_{AB}$ vanishes if and only if  $\rho_{AB}$ has no correlations, i.e., $\rho_{AB}=\rho_{A}\bigotimes \rho_{B}$. The `only if' part is directly derived from the theorem 2 below. In this sense, quantum correlated coherence can be seen as the basis-dependent measure of quantumness of  correlation  and accounts for quantum correlations, for example, quantum discord.

With respect to the fixed  bases of $\{|i\rangle _{A}\}$ and $\{|j\rangle _{B}\}$ of subsystems $A$ and $B$ respectively, the local von Neumann measurements of Alice and Bob are denoted by $\mathrm{\Pi}_{A}= \{|i\rangle _{A}\langle i|\}$ and $\mathrm{\Pi}_{B}= \{|j\rangle _{B}\langle j|\}$, respectively. By direct calculation, we get that the consumption of quantum correlated coherence for any state $\rho_{AB}$ under $\mathrm{\Pi} _{B}$ coincides with the asymmetric basis-dependent discord $D_{\mathrm{\Pi} _{B}}^{A|B}(\rho_{AB})$, i.e., $\mathcal{C}_{re}^{cc}(\rho_{AB})-\mathcal{C}_{re}^{cc}(\mathrm{\Pi}_{B}(\rho_{AB}))=D_{\mathrm{\Pi} _{B}}^{A|B}(\rho_{AB})$. According to  the condition for the asymmetric basis-dependent discord $D_{\mathrm{\Pi} _{B}}^{A|B}(\rho_{AB})$ to vanish [37], we have the following result.

 \textbf{Theorem 1}. Let $\{|i\rangle_{A}\}$ and $\{|j\rangle_{B}\}$ be the fixed bases of subsystems $A$ and $B$, respectively,  and the local von Neumann measurement in the basis $\{|j\rangle_{B}\}$ is denoted as $\mathrm{\Pi}_{B}$. For a given  state $\rho_{AB}$ in system $AB$, its quantum correlated coherence  remains unchanged under $\mathrm{\Pi} _{B}$, i.e., $\mathcal{C}_{re}^{cc}(\rho_{AB})= \mathcal{C}_{re}^{cc}(\mathrm{\Pi}_{B}(\rho_{AB}))$, if and only if there exists a decomposition $\rho_{AB}=\sum_{\alpha}p_{\alpha} \rho_{A}^{\alpha}\bigotimes \rho_{B}^{\alpha}$ such that  $p_{\alpha}$ are  probabilities and all the states $\rho_{B}^{\alpha}$ are perfectly distinguishable by the von Neumann measurement in  the fixed basis $\{|j\rangle _{B}\}$.

In theorem 1, two different states, which are perfectly distinguishable by the von Neumann measurement in the fixed basis $\{|j\rangle _{B}\}$, must have disjoint coherence support. The coherence support of a state is the set of incoherent basis vectors which have  nonzero overlap with the state [37].

Using the very similar arguments as $D_{\mathrm{\Pi} _{B}}^{A|B}(\rho_{AB})$, we  obtain that quantum correlated coherence is corresponding to the symmetric basis-dependent discord $\mathcal{C}_{re}^{cc}(\rho_{AB})=D_{\mathrm{\Pi}_{A}\bigotimes\mathrm{\Pi}_{B}}(\rho_{AB})$. Moreover, we also find the condition for  quantum correlated coherence (the symmetric basis-dependent discord) to vanish.

\textbf{Theorem 2}. Let $\{|i\rangle_{A}\}$ and $\{|j\rangle_{B}\}$ be the fixed bases of subsystems $A$ and $B$, respectively. For a given state $\rho_{AB}$ in system $AB$, its  quantum correlated coherence is equal to zero, i.e., $\mathcal{C}_{re}^{cc}(\rho_{AB})=0$, if and only if there exists a decomposition,
 \begin{equation}
 \rho_{AB}=\sum_{k,l}p_{kl} \rho_{A}^{k}\bigotimes \rho_{B}^{l},
 \end{equation}
 such that $p_{kl}$ are  probabilities, and all the states $\rho_{A}^{k}$ and  $\rho_{B}^{l}$ are perfectly distinguishable by the local von Neumann measurements with respect to the fixed bases $\{|i\rangle _{A}\}$ and $\{|j\rangle _{B}\}$, respectively.

 \textbf{Proof}. To prove the sufficiency,  we will use the following property of von Neumann entropy [36],
 \begin{equation}
 S(\sum_{i}p_{i}\rho_{i})= h(\{p_{i}\})+\sum_{i}p_{i}S(\rho_{i}),
 \end{equation}
where $h(\{p_{i}\})$ is Shannon entropy and all $\rho_{i}$ have support on orthogonal subspaces. Since all  $\rho_{A}^{k}$ and  $\rho_{B}^{l}$ are perfectly distinguishable by the local von Neumann measurements in the fixed bases $\{|i\rangle _{A}\}$ and $\{|j\rangle _{B}\}$, respectively, $\{\rho_{A}^{k(l)}\}$,
$\{\rho_{A}^{k}\bigotimes \rho_{B}^{l}\}$, $\{\rho_{A}^{k(l)diag}\}$, and $\{\rho_{A}^{k diag}\bigotimes \rho_{B}^{l diag}\}$ are sets of states with support on orthogonal subspaces. Direct calculation shows that $\mathcal{C}_{re}^{cc}(\rho_{AB})=0.$

Note that
$$\mathcal{C}_{re}^{cc}(\rho_{AB})=S(\rho_{AB}\|\rho_{A}\bigotimes \rho_{B}) -S\big(\mathrm{\Pi}_{A}\bigotimes \mathrm{\Pi}_{B}(\rho_{AB})\| \mathrm{\Pi}_{A}(\rho_{A})\bigotimes \mathrm{\Pi}_{B}(\rho_{B})\big),$$
where $\mathrm{\Pi}_{A}$ and $\mathrm{\Pi}_{B}$ are the local von Neumann measurements  in the fixed  bases $\{|i\rangle _{A}\}$ and $\{|j\rangle _{B}\}$, respectively. To prove the necessity,  we will use the condition for the quantum relative entropy which is unchanged under a quantum operation [38, 39] and the explicit proof  is presented in the Appendix.

Theorem 2 has several implications. First,  it implies that a state with vanished quantum correlated coherence is a especially classical-classical state [40] but not necessary to be a bipartite incoherent state [8, 13]. Particularly, a  qubit-qubit state with vanished quantum correlated coherence is a  product states or a bipartite incoherent state. More complex cases only  emerge in higher dimension. For example, the following qutrit-qutrit state with vanished quantum correlated coherence, has yet local coherences:
 $$\rho_{AB}=\frac{1}{2}\big(\frac{1}{2}|0\rangle\langle0|+\frac{1}{2}|+_{01}\rangle\langle+_{01}|\big)\bigotimes
  \big(\frac{2}{3}|0\rangle\langle0|+\frac{1}{3}|+_{02}\rangle\langle+_{02}|\big)
   +\frac{1}{2}|2\rangle\langle2|\bigotimes|1\rangle\langle1|, $$
where $|+_{ij}\rangle=\frac{1}{\sqrt{2}}(|i\rangle+|j\rangle)$ and the fixed basis of each subsystem is computable basis. Second, theorem 2 gives the structure of bipartite states  which satisfy the super-additive property of the relative entropy of coherence with equality. This settles an important question left open in previous literature [11, 41] that whether the equality of formula $(3)$ holds if and only if $\rho_{AB}$ is product or incoherent. Finally,
  if we choose the local eigenbases of $\rho_{A}$ and $\rho_{B}$ as the fixed bases of subsystems $A$ and $B$ respectively, these two theorems  reduce to the corresponding results in Ref. [13]. In this sense, our results somewhat generalize the previous results.

The above results show that quantum correlated coherence and the basis-dependent discord are closely related. With equality $\mathcal{C}_{re}^{cc}(\rho_{AB})= D_{\mathrm{\Pi}_{A}\bigotimes\mathrm{\Pi}_{B}}(\rho_{AB})$, the symmetric quantum discord is rewritten as $D(\rho_{AB})= \min _{\{|i\rangle_{A},|j\rangle_{B}\}}\mathcal{C}_{re}^{cc}(\rho_{AB})$. Recall that the symmetric quantum discord based on the pseudo distance of relative entropy can be  represented with the relative entropy of coherence, i.e.,  $\mathcal{C }^{free}(\rho_{AB})= \min_{\{|i\rangle_{A},|j\rangle_{B}\}}\mathcal{C}_{re}(\rho_{AB})$ [12, 28]. Both of the minimization are over all generic bases $\{|i\rangle_{A}\}$ and $\{|j\rangle_{B}\}$ of subsystems $A$ and $B$, respectively. However, one may also consider defining a discord measure $D_{POVM}(\cdot)$ via general local positive-operator-valued measurements (POVMs) on each subsystem [25, 29]. Comparing these three quantifiers of quantum discord, we easily have the inequality $\mathcal{C}^{free}(\rho_{AB})\geq D(\rho_{AB})\geq D_{POVM}(\rho_{AB})$. Whenever these three quantities are zero, the corresponding states are  classical-classical states [26, 40]. Similarly, the asymmetric quantum discord $D^{A|B}(\rho_{AB})$ can also be represented by quantum correlated coherence.

In multipartite systems, the global quantum discord (GQD) [27] can even be rewritten with quantum correlated coherence. It is worth noting that the fixed basis of a multipartite system is the tensor product of the fixed bases of all the subsystems. For a $N$-partite state $\rho_{C_{1}C_{2}\cdots C_{N}}$, its GQD is rewritten as
 $$ D(\rho_{C_{1}C_{2}\cdots C_{N}}) = \min_{\{|i_{1}\rangle_{C_{1}},|i_{2}\rangle_{C_{2}},\cdots,|i_{N}\rangle_{C_{N}}\}}\mathcal{C}_{re}^{cc}(\rho_{C_{1}C_{2}\cdots C_{N}}),$$
 where  the minimization is taken over all generic fixed bases of the $N$-partite system denoted as $\{|i_{1}\rangle_{C_{1}}|i_{2}\rangle_{C_{2}}\cdots|i_{N}\rangle_{C_{N}}\}$,  and with respect to this fixed basis it holds that $\mathcal{C}_{re}^{cc}(\rho_{C_{1}C_{2}\cdots C_{N}})=\mathcal{C}_{re}(\rho_{C_{1}C_{2}\cdots C_{N}})-\sum_{i}\mathcal{C}_{re}(\rho_{C_{i}}).$
With the super-additive property of the relative entropy of coherence as given in forma $(3)$, the GQD is non-negative and for a multipartite classical state it is equal to zero. This provides a simple proof of the non-negativity of GQD in Ref. [27]. These results mean that quantum discord in multipartite systems can be better understood with the framework of quantum coherence.

\section{Quantum correlated coherence and quantum entanglement}
 According to the above discussion, we know that if for arbitrary fixed bases of subsystems $A$ and $B$ the quantum correlated coherence of $\rho_{AB}$ does not vanish there must exist some quantum correlation between subsystems $A$ and $B$, for example, quantum discord. Moreover, it is also possible to characterize entanglement with quantum correlated coherence via state extensions [13], and then entanglement can be seen as the irreducible residue of quantum correlated coherence. This highlights the non-locality of quantum entanglement.

 For a given state $\rho_{AB}$ in system $AB$, a bipartite state $\rho_{AA'BB'}$ is an extension of $\rho_{AB}$ if $\rho_{AA'BB'}$ satisfies $Tr_{A'B'}(\rho_{AA'BB'})=\rho_{AB}$, where subsystems $AA'$ and $BB'$ are held by Alice and Bob, respectively [13, 42]. Via state extensions, the entanglement of $\rho_{AB}$ formulated by quantum correlated coherence is given by  definition 2 below.

  \textbf{Definition 2}. (K. C. Tan et al. [13]) For a given state $\rho_{AB}$, $\rho_{AA'BB'}$ is its unitarily symmetric extension and  let the local eigenbases of  $\rho_{AA'}$ and  $\rho_{BB'}$ be the fixed bases of subsystems $AA'$ and $BB'$, respectively. The entanglement of coherence (EOC) of $\rho_{AB}$ is defined as
  \begin{equation}
 E_{re}^{cc}(\rho_{AB})\equiv \min \mathcal{C}_{re}^{cc}(\rho_{AA'BB'}),
 \end{equation}
 where the minimization is taken over all possible unitarily symmetric extensions  $\rho_{AA'BB'}$.

 In definition 2, the extension $\rho_{AA'BB'}$ is unitarily symmetric if it remains invariant up to local unitary operations on $AA'$ and $BB'$ under a system swap between Alice and Bob. It has been shown that the EOC  has the properties [13]: non-negative and vanished for separated states, invariant under local unitary operations, non-increasing under LOCC operations, and convex. Furthermore, using entropy-based measures, we can even give the bounds of EOC as the following theorem.

\textbf{Theorem 3}. For a given state $\rho_{AB}$, it holds that
\begin{equation}
E_{re}(\rho_{AB})\leq E_{re}^{cc}(\rho_{AB})\leq E_{f}(\rho_{AB}).
\end{equation}
If $\rho_{AB}$ is a pure state, these three quantities in inequality $(8)$ are equal.

 \textbf{Proof}. Taking some unitarily symmetric extension $\rho_{AA'BB'}$ of $\rho_{AB}$, we have
 \begin{equation}
\mathcal{C}_{re}^{cc}(\rho_{AA'BB'})=\mathcal{C}_{re}(\rho_{AA'BB'})\geq E_{re}(\rho_{AA'BB'})\geq E_{re}(\rho_{AB}),
\end{equation}
  where the first inequality  is due to that the relative entropy of coherence is no less than the relative entropy of entanglement for a state [12], and the last inequality is due to that entanglement is un-increased under LOCC operations [21, 22]. With the definition of EOC, formula $(9)$ means that $E_{re}(\rho_{AB})\leq E_{re}^{cc}(\rho_{AB})$.

 To prove the inequality $E_{re}^{cc}(\rho_{AB})\leq E_{f}(\rho_{AB})$, we consider the optimal decomposition of the state $\rho_{AB}=\sum_{i}p_{i}^{*}|\psi_{i}^{*}\rangle\langle\psi_{i}^{*}|$ such that [20]
$E_{f}(\rho_{AB})=\sum_{i}p_{i}^{*}E_{re}(|\psi_{i}^{*}\rangle\langle\psi_{i}^{*}|)$. Every state $|\psi_{i}^{*}\rangle$ is represented with the Schmidt decomposition $|\psi_{i}^{*}\rangle=\sum_{j_{i}}\lambda_{j_{i}}|j_{i}\rangle_{A}|j_{i}\rangle_{B}.$
For every $i$, $\{|j_{i}\rangle_{A}\}$ and $\{|j_{i}\rangle_{B}\}$ are expanded to be the orthonormal bases of subsystems $A$ and $B$, respectively, but both of them are still labeled with original symbols. Define the state
$$\rho_{AA'BB'}^{\triangle}\equiv\sum_{i}p_{i}^{*} \sum_{j_{i},j'_{i}}\lambda_{j_{i}}\lambda_{j'_{i}}|j_{i}\rangle_{A}\langle j'_{i}|\bigotimes
|i\rangle_{A'}\langle i|\bigotimes |j_{i}\rangle_{B}\langle j'_{i}|\bigotimes
|i\rangle_{B'}\langle i|,$$
where $\{|i\rangle_{A'(B')}\}$ is the orthonormal basis of system $A' (B')$. Note that $\{|j_{i}\rangle_{A}|i\rangle_{A'}\}$ and $\{|j_{i}\rangle_{B}|i\rangle_{B'}\}$  are local eigenbases of  $\rho_{AA'}$ and $\rho_{BB'}$, respectively.  Let
$U_{AA'}=\sum_{i,j_{i}}|j_{i}\rangle_{BA}\langle j_{i}|\bigotimes |i\rangle_{B'A'}\langle i|$ and
$U_{BB'}=\sum_{i,j_{i}}|j_{i}\rangle_{AB}\langle j_{i}|\bigotimes |i\rangle_{A'B'}\langle i|$ and a little thought shows that these two unitary operators satisfy $$U_{AA'}\bigotimes U_{BB'} (T_{swap}\rho_{AA'BB'}^{\triangle}T^{\dag}_{swap})U^{\dag}_{AA'}\bigotimes U^{\dag}_{BB'}=\rho_{AA'BB'}^{\triangle},$$ where $T_{swap}$ denotes the swap operator with respect to the local eigenbases of  $\rho_{AA'}$ and $\rho_{BB'}$, i.e., $\{|j_{i}\rangle_{A}|i\rangle_{A'}\}$ and $\{|j_{i}\rangle_{B}|i\rangle_{B'}\}$. Therefore, $\rho_{AA'BB'}$ is unitarily symmetric.
Consequently, we calculate the quantum correlated coherence of $\rho_{AA'BB'}^{\triangle}$,
\begin{eqnarray}
\mathcal{C}_{re}^{cc}(\rho_{AA'BB'}^{\triangle})&=& \mathcal{C}_{re}(\rho_{AA'BB'}^{\triangle})
       =S(\rho_{AA'BB'}^{\triangle diag})-S(\rho_{AA'BB'}^{\triangle})\nonumber\\
   &=& \sum_{i}p_{i}^{*}S(\sum_{j_{i}}\lambda_{j_{i}}^{2}|j_{i}\rangle_{A}\langle j_{i}|\bigotimes|j_{i}\rangle_{B}\langle j_{i}|)\nonumber \\
   &=& \sum_{i}p_{i}^{*}S(\sum_{j_{i}}\lambda_{j_{i}}^{2}|j_{i}\rangle_{A}\langle j_{i}|)=E_{f}(\rho_{AB}),\nonumber
\end{eqnarray}
where the third equality  is due to the property of von Neumann entropy as given by formula $(6)$. The above equality together with  the definition of EOC implies that
$$E_{re}^{cc}(\rho_{AB})\leq \mathcal{C}_{re}^{cc}(\rho_{AA'BB'}^{\triangle})=E_{f}(\rho_{AB}).$$
If $\rho_{AB}$ is a pure state, its relative entropy of entanglement is equal to its entanglement of formation, and then equal to its EOC. Hence, the desired results of theorem 3 are obtained.

   From theorem 3, we conclude that the EOC  is not strictly less than the relative entropy of entanglement for a bipartite state, since for pure states they are equal. Moreover, for a maximally correlated states [15,43], which has the form:
   \begin{equation}
   \rho_{AB}=\sum_{i,j}\rho_{ij} |i\rangle_{A}\langle j|\bigotimes  |i\rangle_{B}\langle j|,
   \end{equation}
   where $\{|i\rangle_{A}\}$ and $\{|j\rangle_{B}\}$ are orthonormal bases of subsystems $A$ and $B$ respectively and $\rho_{ij}$ are the matrix elements,
   its EOC is also equal to its relative entropy of entanglement. We show this result in the following theorem.

\textbf{Theorem 4}. For a  maximally correlated state $\rho_{AB}$ as given by formula $ (10)$, its EOC is equal to its relative entropy of entanglement, i.e., $E_{re}^{cc}(\rho_{AB})=E_{re}(\rho_{AB}).$

 \textbf{Proof}: For the maximally correlated state $\rho_{AB}$, it has the form as given by formula $(10)$. Let the local eigenbases of $\rho_{A}$ and $\rho_{B}$, i.e., $\{|i\rangle_{A}\}$ and $\{|j\rangle_{B}\}$, be the fixed bases of subsystems $A$ and $B$ respectively . According to the Ref. [8], we have
 $\mathcal{C}_{re}(\rho_{A}^{*})= E_{re}(\rho_{AB}),$
where $\rho_{A}^{*}=\sum_{i,j}\rho_{ij} |i\rangle_{A}\langle j|$ in subsystem $A$.
Direct calculation yields
$\mathcal{C}_{re}^{cc}(\rho_{AB})=S(\rho_{AB}^{diag})-S(\rho_{AB})=S(\rho_{A}^{*diag})-S(\rho_{A}^{*})=\mathcal{C}_{re}(\rho_{A}^{*}).$
 Obviously, $\rho_{AB}\bigotimes |00\rangle_{A'B'}\langle 00|$ ia a unitarily symmetric extension of $\rho_{AB}$.  With respect to the local eigenbases of $\rho_{AA'}$ and $\rho_{BB'}$ as the fixed bases of subsystems $AA'$ and $BB'$, respectively, we have the equality
 $\mathcal{C}_{re}^{cc}(\rho_{AB}\bigotimes |00\rangle_{A'B'}\langle 00|) =\mathcal{C}_{re}^{cc}(\rho_{AB}).$
 With the definition of EOC, it holds that
$E_{re}^{cc}(\rho_{AB})\leq \mathcal{C}_{re}^{cc}(\rho_{AB}\bigotimes |00\rangle_{A'B'}\langle 00|)=\mathcal{C}_{re}^{cc}(\rho_{AB}).$
Combining the aforementioned results and theorem 3, we arrive at the result $E_{re}^{cc}(\rho_{AB})=E_{re}(\rho_{AB}).$

   Using the proof of theorem 4, we confirm that for a  maximally correlated state $\rho_{AB}$, its EOC is even equal to its quantum correlated coherence with respect to the  local eigenbases of $\rho_{A}$ and $\rho_{B}$, respectively. Moreover, with theorem 4, it is easy to find a state for which the EOC is strictly less than the entanglement of formation, for example, the maximally correlated Bell diagonal state in the two-qubit system, $\rho_{AB}^{mc}= \frac{3}{4}|\Phi^{+}\rangle \langle\Phi^{+}|+\frac{1}{4}|\Phi^{-}\rangle \langle\Phi^{-}|$, where
 $|\Phi^{\pm}\rangle=\frac{1}{\sqrt{2}}(|00\rangle\pm|11\rangle)$ [20, 21]. However, we do not know whether the EOC  is equal to the relative entropy of entanglement for any mixed state. In addition, for any bipartite state $\rho_{AB}$ and $ \tau_{CD}$, EOC  satisfies the following sub-additivity,
 $$\max \{E_{re}^{cc}(\rho_{AB}),E_{re}^{cc}(\tau_{CD})\}\leq E_{re}^{cc}(\rho_{AB}\bigotimes \tau_{CD})\leq E_{re}^{cc}(\rho_{AB})+ E_{re}^{cc}(\tau_{CD}).$$

    An alternative measure of entanglement formulated by quantum correlated coherence (quantum discord) is defined as $\bar{E}_{re}^{cc}(\rho_{AB})\equiv \min D(\rho_{AA'BB'})$, where the minimization is taken over all possible unitarily symmetric extensions $\rho_{AA'BB'}$ of  $\rho_{AB}$, and
 $D(\rho_{AA'BB'})= \min_{\{|i\rangle_{AA'},|j\rangle_{BB'}\}}\mathcal{C}_{re}^{cc}(\rho_{AA'BB'}),$
 with the minimization over all generic bases $\{|i\rangle_{AA'}\}$ and $\{|j\rangle_{BB'}\}$  of subsystems $AA'$ and $BB'$, respectively. Removing the the property of unitary symmetry of extension $\rho_{AA'BB'}$ in the definition  $\bar{E}_{re}^{cc}(\rho_{AB})$, we denote this new measure of entanglement as $\tilde{E}_{re}^{cc}(\rho_{AB})$. Remarkably, $\tilde{E}_{re}^{cc}(\rho_{AB})$ is equivalent to the  entanglement measure which is the minimal discord over state extensions [35]. Moreover, $\bar{E}_{re}^{cc}(\rho_{AB})$ and  $\tilde{E}_{re}^{cc}(\rho_{AB})$ have the properties: non-negative and vanished for separated states, invariant  under local unitary operations, non-increasing under local operations, convex and upper bounded by entanglement of formation $E_{f}(\rho_{AB})$. However, the properties of $\bar{E}_{re}^{cc}(\rho_{AB})$ and $\tilde{E}_{re}^{cc}(\rho_{AB})$, which are the invariance (non-increasing property) under classical communication  and the relation to the relative entropy of entanglement, are  not clear. In this sense, the EOC  is more advantageous than $\bar{E}_{re}^{cc}(\rho_{AB})$ and $\tilde{E}_{re}^{cc}(\rho_{AB})$.

 In multipartite systems, there exists an entanglement measure like the definition of EOC. For a $N$-partite state $\rho_{C_{1}C_{2}\cdots C_{N}}$,
 $\rho_{C_{1}C'_{1}C_{2}C'_{2}\cdots C_{N} C'_{N}}$ is its unitarily symmetric extension and the local fixed  bases of subsystems are  the eigenbases of  $\rho_{C_{1}C'_{1}}$, $\rho_{C_{2}C'_{2}}$,$\ldots$, and $\rho_{C_{N}C'_{N}}$,  respectively. Then the entanglement of  $\rho_{C_{1}C_{2}\cdots C_{N}}$
 is defined as $ E_{re}^{cc}(\rho_{C_{1}C_{2}\cdots C_{N}})\equiv \min \mathcal{C}_{re}^{cc}(\rho_{C_{1}C'_{1}C_{2}C'_{2}\cdots C_{N} C'_{N}}),$ where the minimization is over all possible unitarily symmetric extensions  $\rho_{C_{1}C'_{1}C_{2}C'_{2}\cdots C_{N} C'_{N}}$ of  $\rho_{C_{1}C_{2}\cdots C_{N}}$,
  $Tr_{C'_{1}C'_{2}\cdots C'_{N}}(\rho_{C_{1}C'_{1}C_{2}C'_{2}\cdots C_{N} C'_{N}})=\rho_{C_{1}C_{2}\cdots C_{N}}$. Note that the extension $\rho_{C_{1}C'_{1}C_{2}C'_{2}\cdots C_{N} C'_{N}}$ is unitarily symmetric if it remains invariant up to local unitary operations on $C_{i}C'_{i}$ and $C_{j}C'_{j}$ under a system swap between $C_{i}C'_{i}$ and $C_{j}C'_{j}$ for any $i,j=1,2,\ldots,N$. Referring to the proofs of EOC as an entanglement measure [13],  we can show that $ E_{re}^{cc}(\rho_{C_{1}C_{2}\cdots C_{N}})$ has the properties: non-negative and vanished for separated states, invariant under local unitary operations, un-increased under LOCC operations, and convex. These results show that the entanglement in multipartite systems can also be formulated by quantum correlated coherence via state extensions.

\section{Conclusions}
In this paper, using entropy-based measures, we have discussed  the concise relationships between quantum coherence and quantum correlations as defined by quantum discord as well as quantum entanglement. The results mean that quantum discord and entanglement can be formulated by quantum correlated coherence. In particular, we gave the condition for quantum correlated coherence (symmetric basis-dependent discord)to vanish, and this condition provides the explicit structure of states which satisfy the super-additive property of the relative entropy of coherence with equality.
 We further proved the lower and upper bounds of EOC  and showed that the EOC is equal to the relative entropy of entanglement in a large number of scenarios including all pure states and maximally correlated states. For pure states, the LOCC monotonicity (monotonicity on average under LOCC operations [24, 44]) of EOC is easily obtained  with theorem 3. However, we do not know whether the EOC of a general mixed state is LOCC  monotone [24, 44], and we leave it open for future research. Finally, we also generalized our results to multipartite settings.

 These results suggest that the quantum properties of correlations originate from the quantum properties of coherence and quantum correlations are better understood with the framework of coherence. We hope that this work is helpful for further understanding quantum correlations and developing quantum technologies.

\begin{acknowledgements}
We thank Jia-Jun Ma for useful comments. This work is supported by the National Natural Science Foundation of China (Grants No. 61572081, No. 61672110, No. 61671082, No. 61601171).
\end{acknowledgements}


\section*{\center Appendix : The explicit structure of states with vanished quantum correlated coherence}

Here we prove that a state $\rho_{AB}$ with vanished quantum correlated coherence has a decomposition as given in formula $(5)$ in the main text.

For a given state  $\rho_{AB}$ with vanished quantum correlated coherence, its symmetric quantum discord is equal to zero, i.e.,  $D(\rho_{AB})=0$. Then,  $\rho_{AB}$ is a classical-classical state [40] with the form
$$\rho_{AB}= \sum_{\alpha,\beta}\lambda_{\alpha\beta}|\psi_{\alpha}\rangle \langle\psi_{\alpha}|\bigotimes|\phi_{\beta}\rangle \langle\phi_{\beta}|,\eqno{(B1)}$$
where  $\{|\psi_{\alpha}\rangle \}$ and $\{|\phi_{\beta}\rangle\}$ are orthonormal bases of subsystems $A$ and $B$, respectively. From the main text, we see that
$$\mathcal{C}_{re}^{cc}(\rho_{AB})=0 \Leftrightarrow  S(\rho_{AB}\|\rho_{A}\bigotimes \rho_{B}) =S\big(\mathrm{\Pi}_{A}\bigotimes \mathrm{\Pi}_{B}(\rho_{AB})\| \mathrm{\Pi}_{A}(\rho_{A})\bigotimes \mathrm{\Pi}_{B}(\rho_{B})\big). \eqno{(B2)}$$
where the von Neumann measurements $\mathrm{\Pi}_{A}$ and $\mathrm{\Pi}_{B}$ are with respect to the fixed bases $\{|i\rangle_{A}\}$ and $\{|j\rangle_{B}\}$  of subsystems $A$ and $B$, respectively.

Recall that the quantum relative entropy is unchanged under a quantum operation $\mathcal{E}$, meaning that $S(\rho\|\sigma)= S(\mathcal{E}(\rho)\|\mathcal{E}(\sigma))$, if and only if there is a recovery operation $\mathcal{R}$ satisfying $\mathcal{R}\circ \mathcal{E}(\rho)=\rho$, $\mathcal{R}\circ \mathcal{E}(\sigma)=\sigma$ [38, 39]. Moreover, there is an explicit formula for the recovery operation:
$\mathcal{R}(X)=\sigma^{\frac{1}{2}}\mathcal{E}^{\dag}\big(\mathcal{E}(\sigma)^{-\frac{1}{2}}X\mathcal{E}(\sigma)^{-\frac{1}{2}}\big)\sigma^{\frac{1}{2}}.$
Here, $\mathcal{E}$ is the local von Neumann measurements $\mathrm{\Pi}_{A}\bigotimes \mathrm{\Pi}_{B}=(\mathrm{\Pi}_{A}\bigotimes \mathrm{\Pi}_{B})^{\dag}$. Applied to formula $(B2)$, the recovery condition says that
$$\begin{array}{crr}
\mathcal{R}(\mathrm{\Pi}_{A}\bigotimes \mathrm{\Pi}_{B}(\rho_{AB}))&=&
\sum_{i,j}\frac{\sum \limits _{\alpha,\beta}\lambda_{\alpha\beta}|\langle i|\psi_{\alpha}\rangle|^{2}|\langle j|\phi_{\beta}\rangle|^{2}}
{\big(\sum \limits _{\alpha,\beta}\lambda_{\alpha\beta}|\langle i|\psi_{\alpha}\rangle|^{2}\big)
\big(\sum \limits _{\alpha,\beta}\lambda_{\alpha\beta}|\langle j|\phi_{\beta}\rangle|^{2}\big)} \\
&&\bigg(\rho_{A}^{\frac{1}{2}}|i\rangle \langle i|\rho_{A}^{\frac{1}{2}}\bigg)
\bigotimes\bigg(\rho_{B}^{\frac{1}{2}}|j\rangle \langle j|\rho_{B}^{\frac{1}{2}}\bigg).
\end{array}\eqno{(B3)}$$
By letting formula $(B3)$ be equal to formula $(B1)$ and pre- and post-multiplying by $\rho_{A}^{-\frac{1}{2}}\bigotimes \rho_{B}^{-\frac{1}{2}}$, we obtain
 $$ \begin{array}{r}
\sum \limits _{i,j}\frac{\sum \limits_{\alpha,\beta}\lambda_{\alpha\beta}|\langle i|\psi_{\alpha}\rangle|^{2}|\langle j|\phi_{\beta}\rangle|^{2}}
{\big(\sum \limits _{\alpha,\beta}\lambda_{\alpha\beta}|\langle i|\psi_{\alpha}\rangle|^{2}\big)
\big(\sum \limits_{\alpha,\beta}\lambda_{\alpha\beta}|\langle j|\phi_{\beta}\rangle|^{2}\big)}|i\rangle \langle i|
\bigotimes|j\rangle \langle j|\\
=\sum \limits _{\alpha,\beta}\frac{\lambda_{\alpha\beta}}{\big(\sum \limits _{\xi}\lambda_{\alpha\xi}\big)\big(\sum \limits _{\gamma}\lambda_{\gamma\beta}\big)}
|\psi_{\alpha}\rangle \langle\psi_{\alpha}|\bigotimes|\phi_{\beta}\rangle \langle\phi_{\beta}|.\nonumber\\
\end{array}\eqno{(B4)}$$
Remarkably,  $\rho_{A}^{-\frac{1}{2}}$ and $\rho_{B}^{-\frac{1}{2}}$ are defined as
$$\rho_{A}^{-\frac{1}{2}}\equiv\sum_{\alpha:\sum\limits_{\beta}\lambda_{\alpha\beta}\neq 0}
\frac{1}{\sqrt{\sum \limits_{\beta}\lambda_{\alpha\beta}}}|\psi_{\alpha}\rangle \langle\psi_{\alpha}|,\quad
\rho_{B}^{-\frac{1}{2}}\equiv\sum_{\beta:\sum\limits_{\alpha}\lambda_{\alpha\beta}\neq 0}
\frac{1}{\sqrt{\sum\limits_{\alpha}\lambda_{\alpha\beta}}}|\phi_{\beta}\rangle \langle\phi_{\beta}|,$$
where all $\sum\limits_{\beta}\lambda_{\alpha\beta}$  and $|\psi_{\alpha}\rangle$  are the eigenvalues and eigenvectors of $\rho_{A}$, and all $\sum\limits_{\alpha}\lambda_{\alpha\beta}$ and $|\phi_{\beta}\rangle$ are the same requirements of $\rho_{B}$. Thus in formula $(B4)$ we exclude terms on either side which  are not in the support of $\rho_{A}$ and $\rho_{B}.$

 Let $i',j',\alpha'$ and $\beta'$ be the values of $i,j,\alpha$ and $\beta$ in formulas $(B3)$ and $(B1)$. After taking the inner product $\langle i'|\langle j'|(\quad )|\psi_{\alpha'}\rangle|\phi_{\beta'}\rangle$ on either side of formula $(B4)$, we have
$$\begin{array}{cll}
\frac{\sum\limits_{\alpha,\beta}\lambda_{\alpha\beta}|\langle i'|\psi_{\alpha}\rangle|^{2}|\langle j'|\phi_{\beta}\rangle|^{2}}{\big(\sum\limits_{\alpha,\beta}\lambda_{\alpha\beta}|\langle i'|\psi_{\alpha}\rangle|^{2}\big)
\big(\sum\limits_{\alpha,\beta}\lambda_{\alpha\beta}|\langle j'|\phi_{\beta}\rangle|^{2}\big)}&
\langle i'|\psi_{\alpha'}\rangle \langle j'|\phi_{\beta'}\rangle &  \\
&=\frac{\lambda_{\alpha'\beta'}}{\big(\sum\limits_{\beta}\lambda_{\alpha'\beta}\big)\big(\sum\limits_{\alpha}\lambda_{\alpha\beta'}\big)}
\langle i'|\psi_{\alpha'}\rangle \langle j'|\phi_{\beta'}\rangle.&
\end {array}\eqno{(B5)}
$$

If $\langle i'|\psi_{\alpha'}\rangle \langle j'|\phi_{\beta'}\rangle\neq 0$, Eq. $(B5)$ means that
$$\frac{\sum\limits_{\alpha,\beta}\lambda_{\alpha\beta}|\langle i'|\psi_{\alpha}\rangle|^{2}|\langle j'|\phi_{\beta}\rangle|^{2}}{\big(\sum\limits_{\alpha,\beta}\lambda_{\alpha\beta}|\langle i'|\psi_{\alpha}\rangle|^{2}\big)
\big(\sum\limits_{\alpha,\beta}\lambda_{\alpha\beta}|\langle j'|\phi_{\beta}\rangle|^{2}\big)}
=\frac{\lambda_{\alpha'\beta'}}{\big(\sum\limits_{\beta}\lambda_{\alpha'\beta}\big)\big(\sum\limits_{\alpha}\lambda_{\alpha\beta'}\big)}
.\eqno{(B6)}$$
Due to Eq. $(B6)$, we confirm that the left sides of $(B6)$ are the same for all $i'$ and $j'$ satisfying  $\langle i'|\psi_{\alpha'}\rangle \langle j'|\phi_{\beta'}\rangle\neq 0$ when we fix $\alpha'$ and $\beta'$. As the same reason,  the right sides of $(B6)$ are the same for all $\alpha'$ and $\beta'$ satisfying  $\langle i'|\psi_{\alpha'}\rangle \langle j'|\phi_{\beta'}\rangle\neq 0$ when we fix  $i'$ and $j'$.

Expanding formula $(B3)$ continuously, we obtain that
$$ \begin{array}{ccl}
\mathcal{R}(\mathrm{\Pi}_{A}\bigotimes \mathrm{\Pi}_{B}(\rho_{AB}))&=&
\sum\limits_{i,j}\frac{\sum\limits_{\alpha,\beta}\lambda_{\alpha\beta}|\langle i|\psi_{\alpha}\rangle|^{2}|\langle j|\phi_{\beta}\rangle|^{2}}{\big(\sum\limits_{\alpha,\beta}\lambda_{\alpha\beta}|\langle i|\psi_{\alpha}\rangle|^{2}\big)
\big(\sum\limits_{\alpha,\beta}\lambda_{\alpha\beta}|\langle j|\phi_{\beta}\rangle|^{2}\big)}\nonumber \\
&&\bigg(\sum\limits_{\alpha} \sqrt{\sum\limits_{\beta}\lambda_{\alpha\beta}}\langle \psi_{\alpha}|i\rangle|\psi_{\alpha}\rangle \bigg)

\bigg(\sum\limits_{\alpha} \sqrt{\sum\limits_{\beta}\lambda_{\alpha\beta}}\langle i|\psi_{\alpha}\rangle\langle \psi_{\alpha}|\bigg)\nonumber \\
&& \bigotimes\bigg(\sum\limits_{\beta} \sqrt{\sum\limits_{\alpha}\lambda_{\alpha\beta}}\langle \phi_{\beta}|j\rangle|\phi_{\beta}\rangle \bigg)
\bigg(\sum\limits_{\beta} \sqrt{\sum\limits_{\alpha}\lambda_{\alpha\beta}}\langle j|\phi_{\beta}\rangle \langle \phi_{\beta}|\bigg). \nonumber\\
\end{array} \eqno{(B7)}$$
By letting  formula $(B7)$ be equal to formula $(B1)$, we firstly consider the case that $|\psi_{\alpha_{1}}\rangle$ $(|\phi_{\beta_{1}}\rangle)$ and all the other $|\psi_{\alpha}\rangle$ $(|\phi_{\beta}\rangle)$ have disjoint coherence support. Without loss of generality, let $\{|i_{1}\rangle, |i_{2}\rangle\}$ and $\{|j_{1}\rangle, |j_{2}\rangle\}$ be the coherence support of $| \psi_{\alpha_{1}}\rangle$ and $ |\phi_{\beta_{1}}\rangle$ respectively.
A litle thought shows that the sum of only the four terms $(i_{1},j_{1})$, $(i_{1},j_{2})$, $(i_{2},j_{1})$ and $(i_{2},j_{2})$ in formula $(B7)$ coincides with the term
$\lambda_{\alpha_{1}\beta_{1}}|\psi_{\alpha_{1}}\rangle\langle\psi_{\alpha_{1}}|\bigotimes|\psi_{\alpha_{1}}\rangle\langle\psi_{\alpha_{1}}|$ in formula $(B1)$.

Secondly, we consider the case that the coherence support of  $|\psi_{\alpha_{1}}\rangle$  has some intersection with that of  other $|\psi_{\alpha}\rangle$, or the coherence support of $|\phi_{\beta_{1}}\rangle$ has some intersection with that of other $|\phi_{\beta}\rangle$. Without loss of generality, let $\{|i_{1}\rangle, |i_{2}\rangle\}$ be the coherence support of $|\psi_{\alpha_{1}}\rangle$ and $|\psi_{\alpha_{2}}\rangle$, and the set of $\{|i_{1}\rangle, |i_{2}\rangle\}$ has no intersection  with the coherence support of other $|\psi_{\alpha}\rangle$ except $|\psi_{\alpha_{1}}\rangle$ and $|\psi_{\alpha_{2}}\rangle$. Similarly, let $\{|j_{1}\rangle, |j_{2}\rangle\}$  be the coherence support of $|\phi_{\beta_{1}}\rangle$ and $|\phi_{\beta_{2}}\rangle$, and the set of $\{|j_{1}\rangle, |j_{2}\rangle\}$ has no intersection  with the coherence support of other $|\phi_{\beta}\rangle$ except $|\phi_{\beta_{1}}\rangle$ and $|\phi_{\beta_{2}}\rangle$. The sum of only the four terms $(i_{1},j_{1})$, $(i_{1},j_{2})$, $(i_{2},j_{1})$ and $(i_{2},j_{2})$ in formula $(B7)$
will  be written as
$$\begin{array}{l}
\sum\limits_{\scriptstyle i=i_{1},i_{2}; \atop \scriptstyle j=j_{1},j_{2}}\frac{\bigg(\sum\limits_{\scriptstyle \alpha=\alpha_{1},\alpha_{2}; \atop \scriptstyle \beta=\beta_{1},\beta_{2}}\lambda_{\alpha\beta}|\langle i|\psi_{\alpha}\rangle|^{2}|\langle j|\phi_{\beta}\rangle|^{2}\bigg)
\big(\sum\limits_{\beta}\lambda_{\alpha_{1}\beta}+\sum\limits_{\beta}\lambda_{\alpha_{2}\beta}\big)
\big(\sum\limits_{\alpha}\lambda_{\alpha\beta_{1}}+\sum\limits_{\alpha}\lambda_{\alpha\beta_{2}}\big)}
{\big(\sum \limits_{\alpha=\alpha_{1},\alpha_{2};\beta} \lambda_{\alpha\beta}|\langle i|\psi_{\alpha}\rangle|^{2}\big)
\big(\sum \limits_{\alpha;\beta=\beta_{1},\beta_{2}} \lambda_{\alpha\beta}|\langle j|\phi_{\beta}\rangle|^{2}\big)}\nonumber\\

\bigg(\frac{\sum\limits_{\beta}\lambda_{\alpha_{1}\beta}}
{\sum\limits_{\beta}\lambda_{\alpha_{1}\beta}+\sum\limits_{\beta}\lambda_{\alpha_{2}\beta}}
|\langle i|\psi_{\alpha_{1}}\rangle|^{2}|\psi_{\alpha_{1}}\rangle\langle\psi_{\alpha_{1}}|

+\frac{\sum\limits_{\beta}\lambda_{\alpha_{2}\beta}}
{\sum\limits_{\beta}\lambda_{\alpha_{1}\beta}+\sum\limits_{\beta}\lambda_{\alpha_{2}\beta}}
|\langle i|\psi_{\alpha_{2}}\rangle|^{2}|\psi_{\alpha_{2}}\rangle\langle\psi_{\alpha_{2}}|\nonumber\\

+\frac{\sqrt{(\sum\limits_{\beta}\lambda_{\alpha_{1}\beta})(\sum\limits_{\beta}\lambda_{\alpha_{2}\beta})}}
{\sum\limits_{\beta}\lambda_{\alpha_{1}\beta}+\sum\limits_{\beta}\lambda_{\alpha_{2}\beta}}
\langle \psi_{\alpha_{1}}|i\rangle\langle i|\psi_{\alpha_{2}}\rangle|\psi_{\alpha_{1}}\rangle\langle\psi_{\alpha_{2}}|\nonumber\\
+\frac{\sqrt{(\sum\limits_{\beta}\lambda_{\alpha_{1}\beta})(\sum\limits_{\beta}\lambda_{\alpha_{2}\beta})}}
{\sum\limits_{\beta}\lambda_{\alpha_{1}\beta}+\sum\limits_{\beta}\lambda_{\alpha_{2}\beta}}
\langle \psi_{\alpha_{2}}|i\rangle\langle i|\psi_{\alpha_{1}}\rangle|\psi_{\alpha_{2}}\rangle\langle\psi_{\alpha_{1}}|\bigg) \nonumber\\

\bigotimes
\bigg(\frac{\sum\limits_{\alpha}\lambda_{\alpha\beta_{1}}}{\sum\limits_{\alpha}\lambda_{\alpha\beta_{1}}+\sum\limits_{\alpha}\lambda_{\alpha\beta_{2}}}
|\langle j|\phi_{\beta_{1}}\rangle|^{2}|\phi_{\beta_{1}}\rangle\langle\phi_{\beta_{1}}|
+\frac{\sum\limits_{\alpha}\lambda_{\alpha\beta_{2}}}{\sum\limits_{\alpha}\lambda_{\alpha\beta_{1}}+\sum\limits_{\alpha}\lambda_{\alpha\beta_{2}}}
|\langle j|\phi_{\beta_{2}}\rangle|^{2}|\phi_{\beta_{2}}\rangle\langle\phi_{\beta_{2}}|\nonumber\\

+\frac{\sqrt{(\sum\limits_{\alpha}\lambda_{\alpha\beta_{1}})(\sum\limits_{\alpha}\lambda_{\alpha\beta_{2}})}}
{\sum\limits_{\alpha}\lambda_{\alpha\beta_{1}}+\sum\limits_{\alpha}\lambda_{\alpha\beta_{2}}}
\langle \phi_{\beta_{1}}|j\rangle\langle j|\phi_{\beta_{2}}\rangle|\phi_{\beta_{1}}\rangle\langle\phi_{\beta_{2}}|\nonumber\\

+\frac{\sqrt{(\sum\limits_{\alpha}\lambda_{\alpha\beta_{1}})(\sum\limits_{\alpha}\lambda_{\alpha\beta_{2}})}}
{\sum\limits_{\alpha}\lambda_{\alpha\beta_{1}}+\sum\limits_{\alpha}\lambda_{\alpha\beta_{2}}}
\langle \phi_{\beta_{2}}|j\rangle\langle j|\phi_{\beta_{1}}\rangle|\phi_{\beta_{2}}\rangle\langle\phi_{\beta_{1}}|\bigg)
\end{array}. \eqno{(B8)}$$

Using formula $(B6)$, we know that the formulas
$$\frac{\sum \limits_{\scriptstyle \alpha=\alpha_{1},\alpha_{2}; \atop \scriptstyle \beta=\beta_{1},\beta_{2}}\lambda_{\alpha\beta}|\langle i|\psi_{\alpha}\rangle|^{2}|\langle j|\phi_{\beta}\rangle|^{2}}
{\big(\sum \limits_{\alpha=\alpha_{1},\alpha_{2};\beta}\lambda_{\alpha\beta}|\langle i|\psi_{\alpha}\rangle|^{2}\big)
\big(\sum \limits_{\alpha;\beta=\beta_{1},\beta_{2}}\lambda_{\alpha\beta}|\langle j|\phi_{\beta}\rangle|^{2}\big)},$$
 are the same for any $i=i_{1},i_{2}$ and $j=j_{1},j_{2}$. Then using the orthonormality of states in sets $\{|\psi_{\alpha_{1}}\rangle, |\psi_{\alpha_{2}}\rangle\}$ and $\{|\phi_{\beta_{1}}\rangle,|\phi_{\beta_{2}}\rangle\}$, and removing the cross terms that contain $|\psi_{\alpha_{1}}\rangle\langle\psi_{\alpha_{2}}|,|\psi_{\alpha_{2}}\rangle\langle\psi_{\alpha_{1}}|,|\phi_{\beta_{1}}\rangle\langle\phi_{\beta_{2}}|$ or
$|\phi_{\beta_{2}}\rangle\langle\phi_{\beta_{1}}|$ in formula $(B8)$, we obtain the simplified form of formula $(B8)$:
$$p_{\alpha_{1}\alpha_{2}\beta_{1}\beta_{2}}(\mu_{1}|\psi_{\alpha_{1}}\rangle\langle\psi_{\alpha_{1}}|
+\mu_{2}|\psi_{\alpha_{2}}\rangle\langle\psi_{\alpha_{2}}|)\bigotimes
(\eta_{1}|\phi_{\beta_{1}}\rangle\langle\phi_{\beta_{1}}|
+\eta_{2}|\phi_{\beta_{2}}\rangle\langle\phi_{\beta_{2}}|),\eqno{(B9)}$$
where $\mu_{1(2)}$, $\eta_{1(2)}$ and $p_{\alpha_{1}\alpha_{2}\beta_{1}\beta_{2}}$  are non-negative, and satisfy

$$\mu_{1(2)}=\frac{\sum\limits_{\beta}\lambda_{\alpha_{1(2)}\beta}}{\sum\limits_{\beta}\lambda_{\alpha_{1}\beta}+\sum\limits_{\beta}\lambda_{\alpha_{2}\beta}},\quad             \eta_{1(2)}=\frac{\sum\limits_{\alpha}\lambda_{\alpha\beta_{1(2)}}}{\sum\limits_{\alpha}\lambda_{\alpha\beta_{1}}+\sum\limits_{\alpha}\lambda_{\alpha\beta_{2}}},$$
 $$ \mu_{1}+\mu_{2}=1,\quad \eta_{1}+\eta_{2}=1,\quad p_{\alpha_{1}\alpha_{2}\beta_{1}\beta_{2}}=\sum_{\scriptstyle \alpha=\alpha_{1},\alpha_{2}; \atop \scriptstyle \beta=\beta_{1},\beta_{2}}\lambda_{\alpha\beta}.$$
What's more, formula $(B9)$ coincides with the sum of partial terms in formula $(B1)$:
$\sum_{\scriptstyle \alpha=\alpha_{1},\alpha_{2}; \atop \scriptstyle \beta=\beta_{1},\beta_{2}}\lambda_{\alpha\beta}|\psi_{\alpha}\rangle \langle\psi_{\alpha}|\bigotimes|\phi_{\beta}\rangle \langle\phi_{\beta}|.$
Finally, other cases that there exist some intersection of coherence support of $|\psi_{\alpha}\rangle$
or $|\phi_{\beta}\rangle$ can be discussed similarly, and the results like formula $(B9)$ will be obtained. Hence, the equality of formula $(B1)$ and $(B7)$ means that $\rho_{AB}$ has a decomposition as given in formula $(5)$ in the main text.



\end{document}